# Markov vs. nonMarkovian processes

# A comment on the paper 'Stochastic feedback, nonlinear families of Markov processes, and nonlinear Fokker-Planck equations' by T.D. Frank

Joseph L. McCauley[+]

Physics Department
University of Houston
Houston, Tx. 77204-5005
jmccauley@uh.edu

[+]Senior Fellow
COBERA
Department of Economics
J.E.Cairnes Graduate School of Business and Public Policy
NUI Galway, Ireland

## Abstract

The purpose of this comment is to correct mistaken assumptions and claims made in the paper "Stochastic feedback, nonlinear families of Markov processes, and nonlinear Fokker-Planck equations" by T. D. Frank [1]. Our comment centers on the claims of a "nonlinear Markov process" and a "nonlinear Fokker-Planck equation". First, memory in transition densities is misidentified as a Markov process. Second, the paper assumes that one can derive a Fokker-Planck equation from a Chapman-Kolmogorov equation, but no proof was offered that a Chapman-Kolmogorov equation exists for the memory-dependent processes considered. A "nonlinear Markov process" is claimed on the basis of a nonlinear diffusion pde for a *1-point* probability density. We show that, regardless of which initial value problem one may solve for the 1-point density, the resulting stochastic process, defined necessarily by the *conditional* probabilities (the transition probabilities), is either an ordinary *linearly generated* Markovian one, or else is a *linearly generated* nonMarkovian process with memory. We provide explicit examples of diffusion coefficients that reflect both the Markovian and the memory-dependent cases. So there is *neither* a "nonlinear Markov process", *nor* a "nonlinear Fokker-Planck equation" for a conditional probability density. The confusion rampant in the literature arises in part from labeling a nonlinear diffusion equation for a 1-point probability density as 'nonlinear Fokker-Planck', whereas neither a 1-point density nor an equation of motion for a 1-point density can define a stochastic process. In a closely related context, we point out that Borland misidentified a translation invariant 1-point probability density derived from a nonlinear diffusion equation as a conditional probability density. Finally, in the Appendix we present the theory of Fokker-Planck pdes and Chapman-Kolmogorov eqns. for stochastic processes with finite memory.

Let $f_n(x_n,t_n; \ldots;x_1,t_1)$ denote the n-point density of a stochastic process $x(t)$, where we have a sequence $(x_n,\ldots,x_1)=(x(t_n),\ldots,x(t_1))$ with $t_n \geq t_{n-1} \geq \ldots \geq t_1$, and let $p_n(x_n,t_n:x_{n-1}t_{n-1};\ldots;x_1t_1)$ denote the corresponding 2-point conditional probability density, the 2-point transition density, depending on history $(x_{n-2},t_{n-2};\ldots;x_1,t_1)$, so that $f_n=p_n f_{n-1}$ [2]. By a Markov process [2,3] is meant that $p_n=p_2(x_n,t_n:x_{n-1},t_{n-1})$ for all $n \geq 3$. That is, there is no history in the transition densities, meaning that $p_n=p_2$ depends only on the last observed point $(x_{n-1},t_{n-1})$ *and on no other history* (see also Remark (9.2.7) on pg. 148 of [4]). In particular, the transition density $p_2$ of a Markov process is entirely independent of initial conditions. This means that the drift and diffusion coefficients depend on a *single point* (x,t) alone, *and on no other states from the past*: e.g., the diffusion coefficient is defined by

$$D(x,t) \approx \frac{1}{T} \int (y-x)^2 p_2(y,t:x,t-T)dy \qquad (1)$$

as T vanishes, and is independent of any initial state $(x_1,t_1)$.

In ref. [1] Frank begins instead with the assumption of transition densities $p_2(x,t:y,s;u)$ that are assumed to depend on the initial condition $p_1(x,t_1)=u(x)$ at the initial time $t_1$, and by $p_1(x,t)$ we mean $f_1(x,t)$. *Although the fact is disclaimed in [1], this is clearly a nonMarkovian assumption*. As is shown explicitly in ref. [5], any dependence of the transition densities on the initial condition is nonMarkovian:

$$p_2(x_3,t_3:x_2,t_2) = \frac{\int p_3(x_3,t_3:x_2,t_2;x_1,t_1)p_2(x_2,t_2:x_1,t_1)p_1(x_1,t_1)dx_1}{\int p_2(x_2,t_2:x_1,t_1)p_1(x_1,t_1)dx_1}$$

.

(2)

That is, with memory, we obtain a functional dependence of the 2-point density $p_2$ on the 1-point density $p_1(x_1,t_1)$ at times $t>t_1$. E.g., with $p_1(x_1,t_1)=\delta(x_1-x'_o)$ at $t_1=0$, we obtain memory $p_2(x_3,t_3:x_2,t_2)=p_3(x_3,t_3:x_2,t_2;x'_o,0)$, necessarily requiring a dependence on $x_o'$ in both the drift and diffusion coefficients and violating the condition for a Markov process. With initial state memory the Markov condition that $p_n=p_2$ must be independent of history for all $n \geq 3$ is impossible, but exactly the opposite claim can be found in [1] (see eqn. (16) in [1] and the statement that follows it). Linear examples of systems with memory of the initial state have been discussed extensively in the literature (see ref. [5] and references therein), and the formalism discussed in [1] for dependence on history does not differ from the formalism of [5]. Through section 3.1 of [1] there is no assumption of nonlinear diffusion in the discussion.

There are several assumptions made uncritically, and presented without analysis, in [1] that require comment. The first is the assumption, without proof and without any justification whatsoever, of the existence of a Chapman-Kolmogorov (section 3.1.2 in [1]), or more accurately denoted Smoluchowski-Markov-Chapman-Kolmogorov (SMCK) eqn.,

$$p_2(x_3,t_3:x_1,t_1;u) = \int p_2(x_3,t_3:x_2,t_2;u)p_2(x_2,t_2:x_1,t_1;u)dx_2 \quad (3)$$

for a process with memory. Although SMCK equations have been proven to exist for at least two specific nonMarkovian systems [6,7], eqn. (3) is not a harmless, easily satisfied assumption and should not merely be assumed to hold without proof for the systems under consideration in [1]. In any case the transition density must satisfy the initial condition

$$p_2(x,t:y,t) = \delta(x-y). \quad (4)$$

With no memory in the transition densities then (3) must hold, whereas with memory the existence of eqn. (3) is not proven. When (3) is not guaranteed then a Kramers-Moyal expansion cannot be used to derive a Fokker-Planck pde, but that is exactly what is done in section 3.1.2 of [1] where the SMCK eqn. (3) is assumed without discussion in the face of memory. We show in the Appendix how a Fokker-Planck pde for a transition density with memory can be obtained without assuming eqn. (3) a priori (see also Friedman [8] for the corresponding derivation of a Fokker-Planck pde for an Ito process in the Markov case). We can identify the processes of [1] as Ito processes (Frank assumes a Langevin description, e.g.), so the assumption in [1] of a linear Fokker-Planck pde can be justified even if eqn. (3) is not assumed or does not hold. The other side of the coin is, however, if one finds transition densities that obey eqn. (3), then that is not evidence for a Markov process if the transition densities are memory dependent, as they are in the Shimizu-Yamada model discussed in part 3.5 of [1]. In fact, that model provides us with a good example of a nonMarkovian process with memory obeying both Fokker-Planck pde and eqn. (3). We will explain below that what Frank has mislabeled as "a nonlinear family of Markov diffusion process" is in the case of the Shimuzi-Yamada model a nonMarkovian Fokker-Planck pde for a transition density with initial state memory described by eqn. (2).

In references [5,9] the memory kernel and nonWiener noise of the Zwanzig-Mori formalism motivated the discussion, and the emphasis there is on obtaining a diffusive pde to describe the 1-point density for systems with memory (however, nonstationary Gaussian transition densities that are much more general than the Zwanzig-Mori class are also considered in [5]). Here is the point: there's an important lesson in that work that foreshadows our criticism here. As is shown for fBm in [10], and for a class of stationary nonMarkov Gaussian processes in [9], *a Fokker-Planck type pde may describe the time evolution of $p_1$ even for nonMarkovian processes*. E.g., the 1-point density of fBm is *exactly* the same as for a Markov process with Hurst exponent H, whereas the transition density $p_2$ for fBm satisfies no such pde and is as far from Markovian as is mathematically possible [10,11]. In particular, a drift-free Markov process has vanishing increment autocorrelations whereas the 2-point transition density for fBm directly reflects arbitrarily long time increment autocorrelations [10,11], *yet both processes have exactly the same 1-point density* that obey exactly the same diffusive pde. With vanishing drift coefficient the class of stochastic processes considered in [1] is restricted to that of vanishing increment correlations, <(x(t)-x(t-T))(x(t+T)x(t))>=0, meaning that memory in conditional probability densities can first appear at the level of $p_3$ [11]. That the transition density for fBm, e.g., fails to satisfy (3) can be proven directly from the conditions stated in Appendix B of [5], where general nonstationary nonMarkovian Gaussian processes are discussed.

We restrict ourselves in what follows to stochastic processes with vanishing drift, R=0, so that all of the action occurs in the diffusion coefficient. This assumption is not necessary for any reason other than to keep the mathematics as simple as is possible [11], and we will generalize to include nonvanishing drift near the end of the paper. Now for the second main point.

In ref. [1], two separate ideas are presented and then mixed together as if they would be the same. First is the assumption of a SMCK equation for a transition density $p_2$ with memory of initial conditions. We will illustrate below how this assumption differs from the question of nonlinear diffusion. I.e., after his implicit assumption of memory in transition densities Frank assumes suddenly that the drift and diffusion coefficients in a Fokker-Planck pde may depend not just on the initially prepared state $p_1(x,t_1)=u(x)$ as a *functional* as in (2) above, but may depend instead directly on the 1-point density *function* at time t>$t_1$, $p_1(x,t)$. That with memory of an initial condition $p_2(x,t:y,s)$ is a *functional* of $p_1(x,t)$ is inherent in (2), but in section 3.2 [1] an entirely *different* assumption is made, namely, that the *diffusion coefficient* is directly a *function* of $p_1(x,t)$. Given this last assumption, and with no other information, the question of memory remains completely open. So let us now enumerate and analyze the possibilities and see what falls out. Is there, or is there not, any deviation from a "normal" Markov process? If so, then where and why? I.e., is there any nontrivial ground for claiming that the formalism of [1] describes a "nonlinear Markov process"?

In Frank's 1-point pde

$$\frac{\partial p_1(x,t)}{\partial t} = \frac{1}{2}\frac{\partial^2}{\partial x^2}(D_p(x,t;p_1(x,t))p_1(x,t)), \qquad (5)$$

clearly, one must first specify a model diffusion coefficient $D_p$ and then solve an initial value problem for the nonlinear pde (5) to obtain $p_1$. We have available as a simple example Borland's model pde [12] for nonlinear diffusion defined by $D_p(p_1)=p_1^{1-q}$. Frank speculates in his conclusion [1] that Borland's model may be 'more nonlinear' than the general nonautonomous case (5), without defining what he means, but we will show this not to be the case at all. Simply consider the space and time translationally invariant case of (5) where $D_p$ depends on (x,t) through $p_1$ alone. Borland's model fits perfectly into Frank's formalism and will in fact prove to be adequately instructive for our purposes. However, with or without Borland's model, we can make a strong prediction at this point: *the sole role of the nonlinear diffusion pde (5) is to provide a model 1-point density $p_1(x,t)$ that defines a specific model diffusion coefficient via $D_p=D(x,t)$.* All nonlinearity stops with (5): the density $p_1(x,t)$ is completely independent of $p_n$, n≥2, because $p_1$ is to be provided by solving (5), which is independent of $p_2$. So given first $D_p$ and then finding a solution $p_1(x,t)$ (if at all possible!) and thereby obtaining $D(x,t)$, the Fokker-Planck pde for $p_2$ is linear,

$$\frac{\partial p_2(x,t:y,s)}{\partial t} = \frac{1}{2}\frac{\partial^2}{\partial x^2}(D(x,t)p_2(x,t:y,s)), \qquad (6)$$

*and it is (6), not (5), that tells us what class of stochastic process is under consideration*.

**To be specific, there are exactly two possibilities, enumerated neither in [1] nor in [12]:** (i) either $p_2$ is independent of the initial condition u(x) for $p_1$, in which case we must obtain a garden variety Markov process linearly generated via (6). Or, (ii) D(x,t) and therefore $p_2$ depends on the initial condition $p_1(x,t_1)=u(x)$, in which case we must obtain a garden variety *linearly generated* martingale with memory via (6). We will illustrate both of these possibilities below using Borland's model, and in neither case is there anything 'nonlinear' about either the Markov process or the martingale with memory. *Kurz gesagt*, a 1-point nonlinear pde (5), taken alone, describes no definite stochastic process, and when the diffusion coefficient determined by the solution $p_1$ (if a solution exists, is positive, etc.) is combined with a Fokker-Planck pde for $p_2$ then we cannot generate anything other than a standard stochastic process, no new theory is required to handle this set of affairs. *This is not the viewpoint expressed in [1]*. The aim of the rest of this comment is to clarify our analysis via explicit examples and, we hope, to eliminate the ground for all of the confusion in the existing literature based on claims of "nonlinear Fokker-Planck pdes" and "nonlinear Markov processes".

We can illustrate our assertions via direct calculations. E.g., if $p_1$ is taken to be the student-t like density derived self-consistently by Borland [12] for the initial state $(x_1,t_1)=(0,0)$, then one can easily check that her solution is normalizable for all t and satisfies the initial condition $p_1(x,0)=\delta(x)$. We next simply substitute that result for $p_1$ into $D_p= p_1^{1-q}$ to define a diffusion coefficient D(x,t) dependent only on the variables (x,t). Now, with $(x_1,t_1)=(0,0)$, Borland's diffusion coefficient belongs to the very general memory free quadratic class $D(x,t)=t^{2H-1}(1+\varepsilon x^2/t^H)$ where H is the Hurst exponent. In her model $H=1/(3-q)$ and $\varepsilon=(q-1)/C^2(q)$, where $C(q)$ is a normalization factor [13]. In this case the linear pde (6) with the initial condition (4), the Fokker-Planck pde or Kolmogorov's second pde, defines a Markov process: the Markov process so-generated is of the standard textbook variety [2,3], nonlinearity of (5) changes nothing. If we restrict to the initial condition $p_2(x,0:0,0)=\delta(x)$, so that we obtain $p_2(x,t:0,0)=p_1(x,t)$,

$$\frac{\partial p_1(x,t)}{\partial t} = \frac{1}{2}\frac{\partial^2}{\partial x^2}(D(x,t)p_1(x,t)), \qquad (6b)$$

then we already know [13] that for the general class of quadratic diffusion coefficients, $D(x,t)=t^{2H-1}(1+\varepsilon x^2/t^H)$, we obtain a 2-parameter $(\varepsilon,H)$ class of student-t like densities. *This is because, for quadratic diffusion and only for quadratic diffusion, the solution $p_1$ of (6) is exactly a power of $D(x,t)$* [13]. For the choice $H=1/(3-q)$ we reproduce precisely Borland's solution from that class. We stated in [13] that Borland's density and diffusion coefficient are completely explained by linear diffusion (6b), and from this standpoint the pde (5) for her specific solution is only superficially nonlinear, is only a linear pde in nonlinear disguise. *But this is not our main point:* **the main point of this paper is that no matter** *how* **one solves or interprets (5), or for** *what* **initial condition one solves (5), once $D_p=D(x,t)$ is obtained then the** *linear* **pde (6) generates the transition probability density for either a Markov process (the case of no memory in $D(x,t)$), or else it linearly generates a nonMarkovian process (the case of memory of u in D), eliminating all ground for the claim in [1] of 'nonlinear Markov processes'**. We will illustrate below a solution of (5) that produces initial state memory in the model defined by $D_p=p_1^{1-q}$. First, we emphasize further that a nonlinear diffusion pde like (5) does not and *cannot* define a stochastic process, cannot tell us whether a process $x(t)$ is or is not Markovian. This means that the criticism of [12] stated in ref. [13] can be seen as incomplete. This comment eliminates that incompleteness.

First, there is no reason to assume uniqueness of solutions of a nonlinear diffusion pde (5), so there may be other solutions of (5) than Borland's self-consistent one, but to date no one has been able to produce one (and, we don't know a priori that such solutions are unique and positive even if they would exist). ***This is also irrelevant for our discussion.*** But the following is relevant. As has been pointed out later [10], and also earlier [9], *a 1-point density $p_1$ cannot be used to decide what sort of stochastic process x(t) is under consideration*. Both Markov and highly nonMarkovian processes like fBm, e.g., can generate exactly the same 1-point density [10], and therefore satisfy exactly the same 1-point diffusion pde, although fBm at the level of $p_2$ is very strongly nonMarkovian [10,11]. The point is that (5), without a specific functional prediction for $D_p=D(x,t)$, does not tell us if a process is Markovian or not. How and whether $p_1$ enters into the definition of $D_p$ is irrelevant so far as classification of stochastic processes goes.

Regardless of whether the process under consideration is or is not Markovian, regardless of whether a SMCK equation (3) holds or not, and independently of (5), we must always require that

$$p_1(x,t) = \int f_2(x,t;y,s)dy = \int p_2(x,t:y,s)p_1(y,s)dy, \quad (7)$$

simply by definition of $p_1$ and $p_2$ [2,3]. Although the pde for $p_1$ may be nonlinear (5), once $p_1$ is known and $D(x,t)$ is determined, e.g. via $D_p=p_1^{1-q}$, then the pde for the transition density $p_2$ (6) is always linear and satisfies

$$0 = (\frac{\partial}{\partial t} - \frac{1}{2}\frac{\partial^2}{\partial x^2}D(p_1(x,t)p_1(x,t)) = \int(\frac{\partial}{\partial t} - \frac{1}{2}\frac{\partial^2}{\partial x^2}D(p_1(x,t)p_2(x,t;y,s))p_1(y,s)dy$$

(8)

Indeed, according to Frank's formalism (see section 3.2 of [1]), given $p_1(x,t)$ as the solution of (5), then $p_2$ must satisfy the linear pde

$$\frac{\partial p_2(x,t;y,s)}{\partial t} = \frac{1}{2}\frac{\partial^2}{\partial x^2}(D_p(p_1(x,t))p_2(x,t;y,s)) \qquad (9)$$

with the Green function initial condition (4). With $D_p=D(x,t)$ this is simply the usual Fokker-Planck pde (6) above, no matter which solution of (5) is used to define $D(x,t)$. *With Borland's solution, or with any other 1-point density, the pde (9) always defines linear diffusion: a Fokker-Planck equation, Kolmogorov's second equation, is defined [2,3] as the equation of motion for the conditional probability density $p_2$ and is always linear.* Another way to state it is that a Langevin equation *necessarily* generates a linear diffusive pde for the transition probability density [13], the Fokker-Planck pde (6).

The way that eqn. (8) works for fBm, where $p_2$ satisfies no Markov-like partial differential equation, is instructive but is not presented here (the nonMarkovian terms in the derivatives under the integral sign must integrate to zero in (8) in that case). The point here is that any nonlinearity in the origin of $p_1$ presents us with no difficulty whatsoever, the standard textbook theory of stochastic processes based on linear partial differential equations for transition probabilities handles this case with ease. We don't know if arbitrary initial value problems $p_1(x,t_1)=u(x)$ of (5) exist, but so far as the Fokker-Planck pde (9) is concerned, we can only obtain either (i) ordinary Markovian dynamics if there is no dependence in $p_1(x,t)$ of the initial state $u(x)$, or (ii) otherwise we will obtain a Martingale with memory [11] through the dependence of the diffusion coefficient on the initial condition $u(x)$. In the latter case the process is not 'nonlinear Markovian' but is instead both linear and nonMarkovian. We will now illustrate the latter point explicitly.

The only remaining possible consequence of (5) is memory of the initial condition for $p_1$ in $p_2$, a martingale with memory. We can easily produce examples of nonMarkovian martingales. From Borland's model an example can be constructed by solving the nonlinear pde (5) self-consistently [12] to obtain a density $p_1(x-x_1,t-t_1)$ satisfying $p(x-x_1,0)= \delta(x-x_1)$. This generalization is trivial because solutions of (6) for $(x,t)$ are also solutions if we translate the x and t origins, simply replacing $(x,t)$ by $(x-x_1,t-t_1)$. That is, the pde (5) with Borland's choice of $D_p$ is translation invariant in both x and t, whereas the Fokker-Planck pde (6) with variable diffusion coefficient depending on both x and t is *not* translation invariant (FX data are not translation invariant in either variable either [14,15]). But, in contrast with the assumption made without proof in [12], the solution $p_1(x-x_1,t-t_1)$ *not* a transition density for a stochastic process, (7) and (8) are not satisfied, we don't even know how to solve (9) for $p_2$ for arbitrary initial states $(y,s) \neq (x_1,t_1)$ in this case.

Again, instead of a obtaining 'nonlinear Markov process', a diffusion coefficient with memory of the initial state $(x_1,t_1)$ *linearly* generates a *nonMarkovian* process: with $D=p_1(x-x_1,t-t_1)^{1-q}$ as in [12], then we get a quadratic diffusion coefficient depending on $(x_1,t_1)$, therefore generating via (9) memory of the initial condition in the 2-point conditional density: if we substitute this result for $D(x-x_1;t-t_1)$ into the Fokker-Planck pde (6), then we get a pde for the conditional density $p_2(x,t:y,s)=p_3(x,t:y,s;x_1,t_1)$, via eqn. (3), for a martingale with memory.

Independently of (5), we can create an entire quadratic diffusion class of martingales with memory via the simple ansatz

$$D(x,t;x_1,t_1) = |t-t_1|^{2H-1}(1+\varepsilon(x-x_1)^2/|t-t_1|^H). \quad (10)$$

We can do the same with any nonquadratic scaling [13] or nonscaling diffusion coefficient $D(x,t)$ as well. All that's needed to create memory is a diffusion coefficient that breaks translation invariance in either x or in t, or in both. Again, financial markets require models with broken symmetry in both (x,t) [15]. In Borland's model the Hurst exponent is restricted to $H=1/(3-q)$, whereas in (10) $(\varepsilon,H)$ are independent parameters.

We don't know if the (as yet unknown) transition density $p_2$ following from (6) with (10) satisfies the SMCK eqn. (3). A Fokker-Planck pde can be obtained from an Ito stochastic differential equation (sde) [13] independently of (3). Ito sdes generate Markov processes iff. the diffusion and drift coefficients are history-independent, depend on (x,t) alone. Ito sdes for martingales that may include memory are analyzed in Durrett [16]. The Fokker-Planck pde is not restricted to Markov processes, but can be derived from an Ito sde using Ito's lemma independently of eqn. (3) [13] and is repeated in the Appendix below. There, we also provide a reason to expect that a memory-dependent diffusion coefficient like (10) should yield a Green function of the Fokker-Planck pde (6) that satisfies the SMCK eqn. (3).

Summarizing, nothing 'nonlinear Markovian' can arise from the formalism of [1], nor is any new stochastic theory required to describe the consequences of a 1-point nonlinear diffusion pde (5). There is apparently no nontrivial content in the phrase 'nonlinear Markov process'. We suggest that the confusion arose in the literature in part because of the concentration on 1-point densities instead of on conditional probability densities, and because of the confusion of translationally invariant 1-point densities with transition probability densities [12]. This led to misnaming the nonlinear diffusion eqn. (5) for a 1-point density as 'nonlinear Fokker-Planck', whereas a Fokker-Planck pde defining a stochastic process, in whole (Markov) or in part (martingale with memory) is always a pde for $p_2$, and that pde is necessarily linear.

This is not a technical point or a squabble over definitions, our comment centers on the excessive reliance in the physics and finance literature on $p_1$ for information that a 1-point density simply does not and cannot provide, regardless of its origin (there is a similar misleading reliance in the literature on Hurst exponents [10]). It was assumed that the pde (5) for a 1-point density defines a 'nonlinear' stochastic process [1,12] in spite of the fact that it was pointed out long ago [9] that a eqn. of motion for $p_1$ cannot tell us which stochastic process is under consideration. The best that the pde (5) can do, with the assumption that $D=p_1^{1-q}$, is to provide a specific model for a diffusion coefficient, and then one can forget (5) altogether.

If we introduce a variable drift coefficient $R(x,t)$ then we lose the martingale property [9], but still have a linear pde with memory (simply introduce a drift term in (6)), the case considered in ref. [5,9].

There is another ground for confusion in the literature: Frank shows that the Shimizu-Yamada model, a model with memory in the drift coefficient, satisfies the SMCK eqn. The lore of statistical physics is that eqn. (3) always implies a Markov process (Feller's counterexample is not widely known). We will show in the Appendix why this is not true for models like Shimizu-Yamada, where there is memory of only a finite number of earlier states. Such processes are nonMarkovian but are indeed very close to Markov processes in that they satisfy both eqns. (3) and (6). However, to label them as 'nonlinear Markovian' is a misconception.

Finally, one should be aware that transition densities (including the Green function of (6) for a general initial condition (4)) do *not* scale with Hurst exponent H [10,11,13] even if both D and $p_1$ scale. A nonMarkovian, nonmartingale example is provided by the 2-point density $p_2$ of fBm, which is does not scale with H nor is it time translationally invariant [10]. Time translationally invariant Gaussians with correlations are treated in [5,9].

**Acknowledgement**

The author thanks Gemunu Gunaratne, Kevin Bassler, Enrico Scalas, and Harry Thomas for discussions of stochastic systems with and without memory, and especially thanks Harry Thomas for comments, criticism, and explanation, useful challenges, and corrections of terminology. He also thanks T.D. Frank for email that stimulated a revision of the original ms.

**Appendix: Fokker–Planck pdes and Chapman-Kolmogorov eqns. for nonMarkovian processes with memory of finitely many eqarlier states**

Beginning with an Ito sde,

$$dx = R(x,t;\{x,t\})dt + \sqrt{D(x,t;\{x,t\})}dB, \quad (A1)$$

where {x,t} denotes the memory of a *finite* nr. k of states $(x_k,t_k;....,x_1,t_1)$, consider the time evolution of any dynamical variable A(x) that does not depend explicitly on t (e.g., $A(x)=x^2$). The sde for A is given by Ito's lemma [4]

$$dA = \left(R\frac{\partial A}{\partial x} + \frac{D}{2}\frac{\partial^2 A}{\partial x^2}\right)dt + \frac{\partial A}{\partial x}\sqrt{D(x,t)}dB \quad (A2)$$

With

$$x(t) = x(s) + \int_s^t R(x(q),q;\{x\})dq + \int_s^t \sqrt{D(x(q),q;\{x\})}dB(q) \quad (A3)$$

we form the conditional average

$$\langle A \rangle_t = \int p_n(x,t:y,s;x_k,t_k;...;x_1,t_1)dy \quad (A4)$$

where n=k+2. Then

$$\langle dA \rangle = \left(\left\langle R\frac{\partial A}{\partial x}\right\rangle + \left\langle \frac{D}{2}\frac{\partial^2 A}{\partial x^2}\right\rangle\right)dt \quad (A5)$$

Using <dA>/dt=d<A>$_t$/dt and integrating by parts while ignoring the boundary terms[1], we obtain

$$\int dx A(x)\left[\frac{\partial p_n}{\partial t} + \frac{\partial(Rp_n)}{\partial x} - \frac{1}{2}\frac{\partial^2(Dp_n)}{\partial x^2}\right] = 0, \quad (A6)$$

---

[1] If the transition density has fat tails, then higher moments will diverge. There, one must be more careful with the boundary terms.

so that for an arbitrary dynamical variable A(x) we get the Fokker-Planck pde (Kolmogorov's second eqn.)

$$\frac{\partial p_n}{\partial t} = -\frac{\partial (R p_n)}{\partial x} + \frac{1}{2}\frac{\partial^2 (D p_n)}{\partial x^2}. \quad (A7)$$

for the 2-point transition density depending on a finite history of n-2 previous states.

With memory, instead of a SMCK eqn. (3) one has the hierarchy of eqns. for 2-point transition densities

$$p_{k-1}(x_k,t_k|x_{k-2},t_{k-2};...;x_1,t_1) = \int dx_{k-1} p_k(x_k,t_k|x_{k-1},t_{k-1};...;x_1,t_1) p_{k-1}(x_{k-1},t_{k-1}|x_{k-2},t_{k-2};...;x_1,t_1). \quad (A8)$$

Suppose, as we have assumed above, that the nonMarkovian stochastic process has only a finite chain of memory of length n-2, $(x_{n-2},t_{n-2};...x_1,t_1)$, that for k≥n we have $p_k=p_n$. Then from (A8) we obtain a SMCK eqn. for the 2-point transition density depending on n-2 states in memory,

$$p_n(x_n,t_n|x_{n-1},t_{n-1};...;x_1,t_1) = \int dy\, p_n(x_n,t_n|y,s;x_{n-2},t_{n-2};...;x_1,t_1) p_n(y,s|x_{n-1},t_{n-1};...;x_1,t_1). \quad (A9)$$

In the Shimizu-Yamada model considered by Frank, the SMCK eqn. (A9) holds for n=3. It would seem that eqn. (A9) should hold for the memory process with n=3 generated by (10), but because we don't know how to calculate the transition density that is not proven. Where there's formalism without ilustrative examples, there may always be an unexpected fly in the ointment.

# References

1. T.D. Frank, *Stochastic feedback, nonlinear families of Markov processes, and nonlinear Fokker-Planck equations*, Physica **A331**, 391, 2004.

2. R.L. Stratonovich. *Topics in the Theory of Random Noise*, Gordon & Breach: N.Y., tr. R. A. Silverman, 1963.

3. M.C. Wang & G.E. Uhlenbeck in *Selected Papers on Noise and Stochastic Processes,* ed. N. Wax, Dover: N.Y., 1954.

4. L. Arnold, *Stochastic Differential Equations*, Krieger, Malabar, 1992.

5. P. Hänggi and H. Thomas, *Zeitschr. Für Physik* B26, 85, 1977.

6. W. Feller, *The Annals of Math. Statistics* **30**, No. 4, 1252, 1959.

**7.** M. Courbage & D. Hamdan, *The Annals of Probability* **22**, No. 3, 1662, 1994.

8. A. Friedman, *Stochastic Differential Equations and Applications,* Academic, N.Y., 1975.

9. P. Hänggi, H. Thomas, H. Grabert, and P. Talkner, *J. Stat. Phys*. 18, 155, 1978.

10. J. L. McCauley , G.H. Gunaratne, & K.E. Bassler, *Hurst Exponents, Markov Processes, and Fractional Brownian Motion, Physica **A*** (2007), in press.

11. J. L. McCauley, G.H. Gunaratne, & K.E. Bassler, *Martingales, detrending data, and the efficient market hypothesis*, preprint 2007.

12. L. Borland, Quantitative Finance **2**, 415, 2002.

13. K.E. Bassler, G.H. Gunaratne, & J. L. McCauley, *Physica A* **369**,343 (2006).

14. S. Gallucio, G. Caldarelli, M. Marsilli, and Y.-C. Zhang, *Physica **A245***, 423, 1997.

15. K.E. Bassler, J. L. McCauley, & G.H. Gunaratne, *Nonstationary Increments, Scaling Distributions, and Variable Diffusion Processes in Financial Markets*, 2006.

16. R. Durrett, *Brownian Motion and Martingales in Analysis*, Wadsworth, Belmont, 1984.